\pdfoutput=1

\documentclass[11pt]{article}

\usepackage[final]{acl}

\usepackage{times}
\usepackage{latexsym}
\usepackage{multirow} 
\usepackage{booktabs}  
\usepackage{amsmath}
\usepackage{amsfonts}
\usepackage{amssymb}
\usepackage[T1]{fontenc}

\usepackage[utf8]{inputenc}

\usepackage{microtype}

\usepackage{inconsolata}

\usepackage{graphicx}
\usepackage{cleveref}
\usepackage{threeparttable}
\title{In-the-wild Audio Spatialization with Flexible Text-guided Localization}

\author{Tianrui Pan, Jie Liu\thanks{Corresponding author (liujie@nju.edu.cn
).}, Zewen Huang, Jie Tang, Gangshan Wu \\
        State Key Laboratory for Novel Software Technology, Nanjing University \\
        {a24164839@163.com}, {liujie@nju.edu.cn}, \\{502023370017@smail.nju.edu.cn}, \{tangjie,gswu\}@nju.edu.cn \\
}

\begin{document}
\maketitle
\begin{abstract}
To enhance immersive experiences, binaural audio offers spatial awareness of sounding objects in AR, VR, and embodied AI applications. While existing audio spatialization methods can generally map any available monaural audio to binaural audio signals, they often lack the flexible and interactive control needed in complex multi-object user-interactive environments. To address this, we propose a Text-guided Audio Spatialization (TAS) framework that utilizes flexible text prompts and evaluates our model from unified generation and comprehension perspectives. Due to the limited availability of premium and large-scale stereo data, we construct the SpatialTAS dataset, which encompasses 376,000 simulated binaural audio samples to facilitate the training of our model. Our model learns binaural differences guided by 3D spatial location and relative position prompts, augmented by flipped-channel audio. It outperforms existing methods on both simulated and real-recorded datasets, demonstrating superior generalization and accuracy. Besides, we develop an assessment model based on Llama-3.1-8B, which evaluates the spatial semantic coherence between our generated binaural audio and text prompts through a spatial reasoning task. Results demonstrate that text prompts provide flexible and interactive control to generate binaural audio with excellent quality and semantic consistency in spatial locations. Dataset is available at \href{https://github.com/Alice01010101/TASU}{https://github.com/Alice01010101/TASU}.
\end{abstract}

\section{Introduction}

Humans can identify the location of objects by processing auditory differences between their ears, even when they cannot see or are not physically present in the scene. Binaural audio contains spatial information for each sound source, it is essential for applications in Virtual Reality (VR) or Augmented Reality (AR)~\citep{li2018scene, kim2019immersive, xu2024sounding}, and embodied AI~\citep{liu2024aligning}. The audio spatialization task~\citep{gao20192, zhou2020sep, rachavarapu2021localize, parida2022beyond, garg2023visually, dagli2024see} continues to be a vibrant area of research. This task involves mapping monaural audio signals to binaural audio signals, allowing users to experience immersive surroundings as if they were physically present in the scenes. Most existing methods are visually guided~\citep{gao20192,zhou2020sep,garg2023visually}, performing mono-to-binaural mapping using visual frames captured by cameras of different Field Of Views (FOV). However, accurate mapping between sound sources in binaural audio and visible objects in frames is impeded by sound sources located outside the camera's view and complex environments with extraneous noise. 

\begin{figure}[t]
    \centering
    \includegraphics[width=0.5\textwidth]{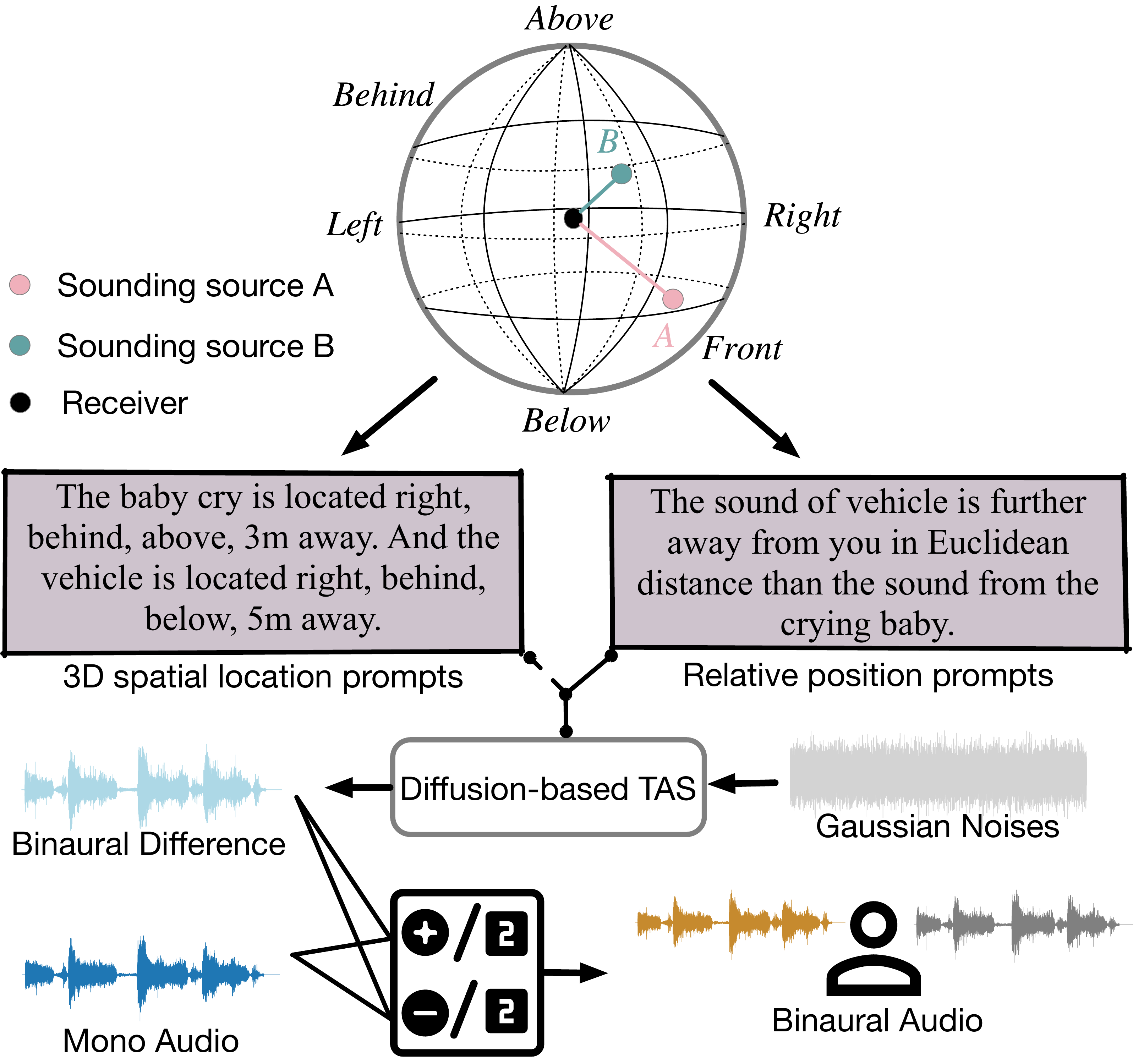}
    \caption{We propose the Text-guided Audio Spatialization (TAS) framework. It utilizes diverse text descriptions to specify the 3D spatial information of multiple sound sources, serving as prompts to transform monaural audio into binaural audio in complex environments.}
    \label{fig:introduction} 
\end{figure}

To address these challenges, we propose a Text-guided Audio Spatialization (TAS) framework that incorporates flexible text prompts and evaluates our model innovatively from the perspective of unified generation and understanding. To the best of our knowledge, the only relevant study in this area \citet{li2024tas} manually labeled text prompts for the FAIR-Play dataset \citep{gao20192} from extracted visual frames, resulting in suboptimal performance due to its simplistic approach and limited dataset scale. To mitigate the lack of corresponding datasets, we propose sampling from a large-scale simulated binaural dataset~\citep{zheng2024bat} and refining it with more detailed text descriptions. This results in the SpatialTAS dataset, which contains approximately 376K training samples. Since providing precise azimuth or elevation information is not always feasible in practical scenarios, we generate two primary types of descriptions, as illustrated in \Cref{fig:introduction}. The first type categorizes eight spatial directions based on the Cartesian product of spherical coordinates: \textit{(left, right)}, \textit{(front, behind)}, \textit{(above, below)}, along with their distances to the receiver. In certain real-time interaction scenarios, humans can only make subjective judgments about the relative spatial relationships between two concurrently active sound events. For the second type, we offer descriptions of the relative positions between any two sound sources. These text descriptions enable selective location instructions for specific target objects, thereby enhancing user-friendliness and adaptability to various contexts. This is in contrast to most previous methods~\citep{gao20192, zhou2020sep, rachavarapu2021localize, parida2022beyond, garg2023visually, dagli2024see} that require guidance for all sound sources within an audio mixture to avoid obvious performance drop. Inspired by PseudoBinaural~\cite{xu2021visually}, we aim to train our model on the constructed large-scale simulated SpatialTAS dataset, which can transfer freely to in-the-wild monaural audios (\Cref{sec:in-the-wild results}).


Recent works~\citep{li2024cyclic, li2024tas} have achieved impressive performance using diffusion models in the audio spatialization task. However, these approaches employ a diffusion model directly in the waveform space, utilizing a cross-attention module to interact with audio and text embeddings. In contrast to this approach, we leverage a latent diffusion model~\citep{rombach2022high} that is directly conditioned on text embeddings to learn the binaural difference between the left and right audio channels, as illustrated in~\Cref{fig:introduction}. By learning the latent representations of audio signals without modeling the cross-modal relationship, our model improves both generation quality and computational efficiency. Furthermore, recognizing the absence of spatial audio alignment in the pretrained text encoder during training, we introduce a text-audio coherence module. This module employs flipped-channel audio to finetune the encoder, thereby enriching the spatial representation of text embeddings.

While numerous metrics exist for evaluating monaural audio, specific metrics for generated binaural audio remain lacking. In this work, we first establish an assessment model by finetuning Llama-3.1-8B~\citep{dubey2024llama} on the SpatialTAS with the spatial audio reasoning task. Then we utilize the assessment model to assess the spatial semantic coherence between our generated audio and text prompts. Experimental results on the SpatialTAS dataset demonstrate that our generated binaural audio not only exhibits high audio quality but also captures distinct and interpretable spatial characteristics for spatial audio understanding. Furthermore, it shows strong generalization ability when tested on the FAIR-Play~\citep{gao20192} and 360$^{\circ}$ YouTube-Binaural~\citep{garg2023visually} datasets, which consist of real-world binaural recordings, including various audio types such as music, speech, and natural sounds.


    
    

\section{Related Work}
\subsection{Audio Spatialization}
Some studies utilize video frames for self-supervision to infer the positions of sound-emitting objects~\citep{morgado2018self,garg2023visually,gao20192,zhou2020sep}. \citet{morgado2018self} introduced two datasets for audio spatialization using 360$^\circ$ videos: REC-STREET and YT-ALL. \citet{garg2023visually} enhanced the YT-Clean dataset by converting ambisonic audio to binaural audio with Normal Field-Of-View (NFOV) video clips, creating the YouTube-Binaural dataset, which we use alongside the original 360$^\circ$ videos. \citet{gao20192} proposed the FAIR-Play dataset, focusing on NFOV video and binaural audio with multiple music tracks. Other studies improved alignment between binaural audio and visual features\citep{garg2023visually,liu2024visually,li2024cyclic}. Recently, \citet{li2024tas} labeled the FAIR-Play dataset with object location descriptions and suggested guiding audio spatialization with text. 



\begin{table*}[ht]
    \centering
    \footnotesize
    \caption{\textbf{Overview of text condition types.} The SpatialTAS dataset includes about 256,000 samples with 3D spatial location and distance prompts for each sound source, along with approximately 120,000 samples for relative spatial relationships among multiple sound sources. \textbf{Sources} indicates the number of sound sources present in each sample.}
    \label{tab: data_generation}
    \begin{tabular}{ccl}
    \toprule
       \textbf{Text Type} &\textbf{Sources} & \textbf{Example} \\
    \midrule
        \multirow{3}{*}{\shortstack[l]{\textbf{DOA \& DE} \\ (256K, 68\%)}}  & 1 &\textbf{A:} The emergency vehicle is located right, behind, below, 5m away. \\
        & \multirow{2}{*}{2} &\textbf{B:} The music is located left, behind, below, 8.5m away. And the whip is located right, \\
        & \quad &\ behind, below, 5m away.\\
    \midrule 
        \multirow{8}{*}{\shortstack[c]{\textbf{Relative} \\ \textbf{Relationships} \\ (120K, 32\%)}} & \multirow{8}{*}{2} & \textbf{C:} The distance between the sound of the animal and the sound of the spray is 3m away. \\
        && \textbf{D:} The sound from the music on the back is located further away, while the sound from the \\
        && \quad \ telephone dialing with DTMF is closer to the front. \\
        
        && \textbf{E:} The sound from the scratching originates on the left, and the sound from the children \\
        && \quad \ playing originates on the right. \\
        
        && \textbf{F:} The sound from the music is above and the sound from the boat, water vehicle is below. \\

        && \textbf{G:} The sound from speech is further away from you in Euclidean distance than the sound \\
        && \quad \ from a mechanical fan. \\

    \bottomrule
    \end{tabular}
\end{table*}

\subsection{Binaural Audio Generation}
Recently, several text-to-audio generation methods have been proposed~\citep{liu2023audioldm, liu2024audioldm, vyas2023audiobox, evans2024long, evans2024stable, lee2023voiceldmtexttospeechenvironmentalcontext,yang2023uniaudioaudiofoundationmodel}, with some focusing specifically on text-to-binaural audio generation~\citep{singh2024diff, sun2024both}. \citet{singh2024diff} utilized text as the sole input, introducing a multi-conditional encoder to unify spatial and semantic information for context-aligned binaural audio generation. Similarly, \citet{sun2024both} proposed the BEWO-1M dataset, demonstrating a novel approach with promising results. Since large-scale monaural datasets are readily available in the real world, we focus on text-guided audio spatialization, leveraging text prompts to provide flexible and interactive control that better aligns with real-world application needs.

\section{Method}
\subsection{Generating Prompts for Training}
Our object is to establish a text-guided audio spatialization framework that uses positional text descriptions $T_{\text{prompts}}$ to transform a monaural audio $A_{\text{mono}}$ into binaural audio, along with a unified evaluation for generation and understanding.

Given the limited scale of most real-recorded binaural audio datasets and the absence of text prompts for sound source locations, we introduce SpatialTAS, a large-scale simulated dataset meticulously crafted by sampling and refining data from the SpatialSoundQA~\citep{zheng2024bat} dataset by GPT-4o~\citep{hurst2024gpt}. Notably, SpatialTAS incorporates more fine-grained text prompts tailored for binaural audio generation. As detailed in Table~\ref{tab: data_generation}, our dataset encompasses approximately 256K samples with text descriptions for Direction Of Arrival (DOA) and Distance Estimation (DE), complemented by an additional 120K samples featuring descriptions of relative relationships between sound sources. The SpatialTAS dataset provides comprehensive 3D spatial location prompts that convey the direction and distance of each sound source, along with versatile relative position prompts that facilitate flexible specification of spatial relationships between any two sound sources. In Table~\ref{tab: data_generation}, Example \textbf{A} and Example \textbf{B} exemplify detailed spatial location prompts. Example \textbf{A} represents a scenario with a single sounding source, while Example \textbf{B} depicts a situation with two sound sources in an audio mixture. Regarding the versatile relative position prompts, Example \textbf{C} conveys information about the relative distance between the two sources, whereas Examples \textbf{D}, \textbf{E}, \textbf{F}, and \textbf{G} describe their relative spatial locations. The dataset comprises hundreds of diverse audio events carefully selected from 10-second audio clips in AudioSet~\citep{gemmeke2017audio}. We aim to train a model on this large-scale simulated dataset, enabling seamless transfer to in-the-wild audios.

\subsection{Audio Spatialization Framework}
\begin{figure*}[h]
    \centering
    \includegraphics[width=1.0\textwidth]{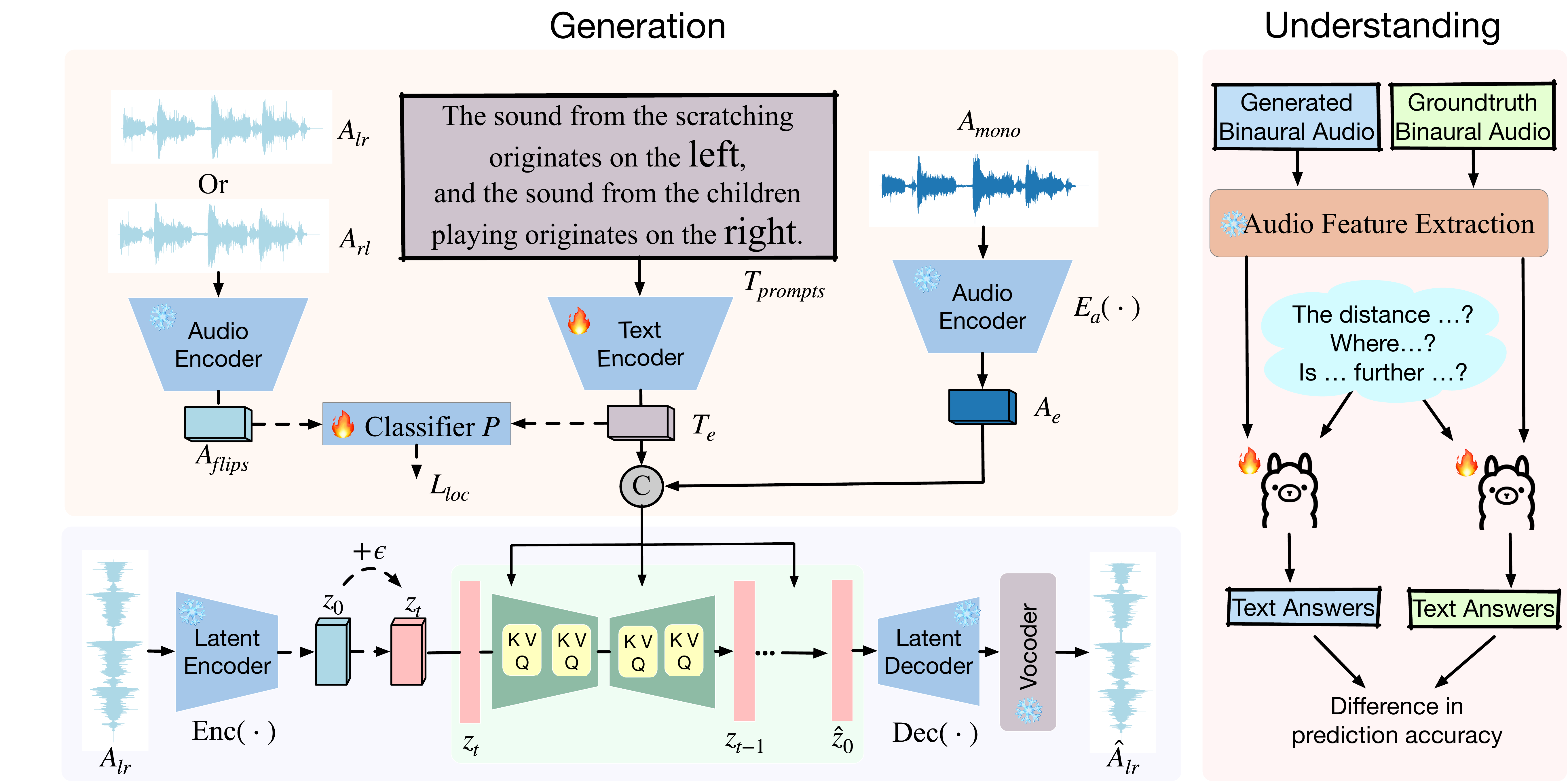}
    \caption{\textbf{The detailed structure for the text-guided audio spatialization model.} The dashed lines indicate processes that occur only during training. We train a latent diffusion model that adds noise to the monaural audio $A_{\text{mono}}$ based on the concatenation of the encoded text embedding $T_{e}$ and audio embedding $A_{e}$. During inference, the model predicts the binaural difference  $A_{lr}$ from the Gaussian noise. Additionally, we finetune a LLM to perform spatial reasoning, verifying the accuracy of the spatial semantic information in our generated binaural audio.}
    \label{fig:method} 
\end{figure*}

During training, we train a diffusion model to learn the channel difference $A_{lr}$ from Gaussian noise, which signifies the distinction between the left and right channels. Given the simulated binaural audio $A_b=(A_l, A_r)$, the monaural audio $A_{\text{mono}}$ is obtained by mixing the left and right channels, while the target channel difference audio $A_{lr}$ is obtained by subtracting the channels:
\begin{equation}
    A_{\text{mono}} = A_l + A_r, A_{lr} = A_l - A_r,
\end{equation}
where $A_{\text{mono}}$ is utilized as model input during training. We train the latent diffusion model to learn the channel difference $A_{lr}$ from the Gaussian distribution. During inference, we compute generated binaural audio $\hat{A_{b}}=(\hat{A}_l, \hat{A}_r)$ as follows:
\begin{equation}
    \hat{A}_l = \frac{A_{\text{mono}} + A_{lr}}{2},
    \hat{A}_r = \frac{A_{\text{mono}} - A_{lr}}{2}.
\end{equation}
The generated binaural audio $\hat{A}_b$ retains the same spatial positional information for each sound source as in $A_b$. Furthermore, the model demonstrates strong generalization capabilities for real-world binaural audio generation, encompassing various audio types, including music, speech, and diverse sound effects. We train a latent diffusion model $F_{\theta}$ to learn the binaural difference $A_{lr}$ conditioned on text embeddings $T_{e}$ and embedded monaural audio $A_{e}$, together with a spatial coherence module. 

\noindent \textbf{Conditional latent diffusion model.} 
As illustrated in the lower left of \Cref{fig:method}, we employ a Variational AutoEncoder (VAE)~\citep{kingma2013auto} latent encoder $\text{Enc}(\cdot)$ to compress the mel-spectrogram of $A_{lr}$, which has a shape of $ \mathbb{R}^{T \times F} $, into a compact continuous representation $z \in \mathbb{R}^{\frac{T}{r} \times \frac{F}{r} \times C}$. Here, $T$ and $F$ represent the time length and frequency dimensions, respectively. $C$ denotes the number of channels in the latent representation, and $r$ is the downsampling ratio that determines the compression level of the latent space. After the diffusion model, we use the VAE latent decoder $\text{Dec}(\cdot)$ to reconstruct the latent representation $z$ back into the mel-spectrogram format of $A_{lr}$. Additionally, we incorporate a HiFi-GAN vocoder~\citep{kong2020hifi} to convert the mel-spectrogram into a high-quality waveform. Both the latent encoder $\text{Enc}(\cdot)$ and the latent decoder $\text{Dec}(\cdot)$ consist of stacked convolutional modules.

Given the encoded latent representation of the input audio $z_0 = \operatorname{Enc}(A_{lr})$, we apply a forward process during training to obtain the noised representation $z_t$ at each time step $t$. This is done by injecting noise $\epsilon$ according to the equation $z_t = \alpha z_0 + \beta \epsilon$, following the noise schedule~\citep{song2020denoising}. Here, $\epsilon$ is random noise drawn from an isotropic Gaussian distribution $\mathcal{N}(0,I)$. We define the training loss $\mathcal{L}_{\theta}$ as the objective to predict the noise $\epsilon$ added to the noisy latent representation, guided by the text embedding $T_{e}$ and the embedding of the monaural audio $A_{e}=E_a(A_{\text{mono}})$. This is achieved by minimizing the following loss function:
\begin{equation}
    \centering
    \begin{aligned}
    \mathcal{L}_{\theta}=
    \mathbb{E}_{\epsilon \in \mathcal{N}(0,I),t}||\epsilon - F_\theta(z_t,t,T_{e}, A_{e})||_2^2.
    \end{aligned}
\end{equation}
The Classifier-Free Guidance (CFG)~\citep{ho2022classifier} is crucial for generating audio that semantically matches and temporally aligns with the text instructions, while preserving the model's generative diversity and enhancing its generalization capability. Therefore, during training, we randomly replace the condition pair $(T_{e}, A_{e})$ with a zero tensor with a probability of 0.1. And during sampling, we modify the vector field using the formula as follows:
\begin{equation}
\label{eq:cfg}
\begin{aligned}
    &\hat{F}_\theta(z_t, t, T_{e}, A_{e})= \\
    &\gamma F_\theta(z_t, t, T_{e}, A_{e}) + (1 - \gamma) F_\theta(z_t, t, \varnothing, \varnothing),
\end{aligned}
\end{equation}
where $\gamma$ is the guidance scale trading off the sample diversity and generation quality, and $\hat{F}_\theta$ degenerates into the original vector field $F_\theta$ when $\gamma=1$.

\noindent \textbf{Text and audio embeddings.} 
CLAP~\citep{elizalde2023clap} and T5~\citep{raffel2020exploring} are commonly used models for extracting text embeddings. While CLAP captures global features, it lacks temporal sensitivity~\citep{elizalde2023clap}. An ablation study by \citet{sun2024both} shows that CLAP accelerates convergence compared to T5 but performs worse in spatial tasks. To improve text embeddings with better temporal cues and spatial information, we utilize the pretrained FLAN-T5 language model~\citep{chung2024scaling}. This enhanced version of T5 has been fine-tuned on a variety of tasks, enabling us to extract text embeddings $T_e$ from $T_{prompts}$.

\noindent \textbf{Text spatial coherence augmentation.} Most audio-language models, such as CLAP~\citep{elizalde2023clap} and FLAN-T5~\citep{chung2024scaling}, lack specialized training on datasets that provide detailed text spatial coherence for sound localization. To address this, we propose a module that enhances the spatial expressive capacity of the text embeddings. We generate misalignment samples between $A_{lr}:=A_{l}-A_{r}$ and the flipped $A_{rl}:=A_{r}-A_{l}$ to capture spatial localization differences. As shown in the upper left of \Cref{fig:method}, the classifier $P$ integrates the selected features with the text representation $T_{e}$ to assess whether the audio differences align with the text descriptions. This encourages the text features to reason about the relative positions of sound sources and identify cues indicating the perceived direction of sound. To evaluate the classifier's performance in predicting audio flipping, we calculate the Binaural Cross-Entropy (BCE) loss, represented as ground truth indicator $g=P(A_{lr} | A_{rl}, T_{e})$, where $|$ denotes the logical $\operatorname{OR}$ operation. The indicator $g$ indicates the ground truth of whether the audio is flipped, leading to the computation of the BCE loss for spatial coherence as follows:

\begin{equation}
    \mathcal{L}_{loc} = \operatorname{BCE}\left(P(A_{lr} | A_{rl}, T_{e}), g\right).
\end{equation}
The total loss is the combination of the diffusion loss $\mathcal{L}_{\theta}$ and the spatial coherence loss $\mathcal{L}_{loc}$. The $\mathcal{L}_{\theta}$ is aimed at optimizing the parameters of the diffusion model, while $\mathcal{L}_{loc}$ is mainly designed to finetune the text encoder. 

\subsection{Spatial Understanding Metrics}
In addition to evaluating audio quality through generation metrics, we assess the spatial semantic coherence between our generated binaural audio and text prompts using a spatial audio reasoning task. This evaluation is detailed in the understanding part of \Cref{fig:method}. Firstly, we follow \citet{zheng2024bat} to develop an assessment model for spatial audio question answering. We fine-tune the Llama-3.1-8B model~\citep{dubey2024llama} on SpatialTAS, integrating the pretrained SpatialAudioEncoder~\citep{zheng2024bat} to extract spatial audio features. Secondly, we send the ground-truth binaural audio and our generated binaural audio to the assessment model along with the corresponding spatial questions, obtaining the prediction accuracy discrepancy between them. A lower discrepancy indicates superior spatial fidelity in our generated binaural audio. Spatial question types are detailed in \Cref{sec:qa-type}. 

\section{Experiments}
\subsection{Model Implementation Details}
\label{sec:implementation_details}
For our experiments, we employ the pretrained VAE and HiFi-GAN vocoder~\citep{kong2020hifi} from \citet{liu2024audioldm}, with both modules frozen during training. It is trained on the combination of AudioSet~\citep{gemmeke2017audio}, AudioCaps~\citep{kim2019audiocaps}, BBC Sound Effects and Freesound~\citep{fonseca2021fsd50k} datasets.
Our model utilizes a U-Net backbone for the diffusion process, consisting of four encoder and decoder blocks that incorporate downsampling and upsampling operations, with a bottleneck layer positioned between them. Multi-head attention is employed in the last three encoder blocks and the first three decoder blocks, featuring 64 head dimensions and 8 heads per layer. The Variational Autoencoder (VAE) is configured with a compression level $r$ of 4 and a latent dimension $d$ of 8. During the forward process, we implement $N$=1000 steps with a linear noise schedule that ranges from $\beta_1$=0.0015 to $\beta_N$=0.0195 for noise generation. Additionally, we leverage the DDIM sampling method~\citep{song2020denoising} with 200 sampling steps. For classifier-free guidance, we set the guidance scale $\lambda$ to 2.5, as detailed in Equation \Cref{eq:cfg}. Training is conducted using the AdamW optimizer~\citep{loshchilov2017decoupled} with a learning rate of $10^{-4}$, $\beta_1$=0.95, $\beta_2$=0.999, $\epsilon$=$10^{-6}$, and a weight decay of $10^{-3}$ training for 500,000 steps.

\subsection{Dataset}
\noindent \textbf{SpatialTAS Dataset.}
The SpatialTAS dataset, derived from the SpatialSoundQA~\citep{zheng2024bat} dataset, contains large-scale simulated binaural audio with detailed and flexible text descriptions of sound source locations. We split the dataset into 376,104 training samples, 732 validation samples, and 4,000 testing samples. The training samples consist of 138,338 single-object DOA and DP samples, 117,519 two-object DOA and DP samples, 50,501 relative direction samples, and 52,747 relative distance samples. The testing samples are evenly distributed, with 1,000 samples for each category.

\begin{table*}
    \centering
    \begin{threeparttable}
    \begin{tabular}{c cccc cccc}
       \toprule
       \multirow{3}{*}{Method}  &  \multicolumn{4}{c}{Generation Metrics} & \multicolumn{4}{c}{Understanding Metrics}\\
       \cline{2-9}
       & \multirow{2}{*}{FD$\downarrow$} & \multirow{2}{*}{FAD$\downarrow$} & \multirow{2}{*}{KL$\downarrow$} & \multirow{2}{*}{IS$\uparrow$} & \multicolumn{2}{c}{Perception} & \multicolumn{2}{c}{Reasoning} \\
       \cline{6-9}
       &&&&&DOA$\downarrow$ & DE$\downarrow$ & Direction$\downarrow$ & Distance$\downarrow$ \\
       \midrule
       \textcolor{gray}{Mono-Mono} & \textcolor{gray}{9.03} & \textcolor{gray}{3.67} & \textcolor{gray}{0.99} & \textcolor{gray}{1.61} & \textcolor{gray}{19.66} & \textcolor{gray}{18.12} & \textcolor{gray}{12.79} & \textcolor{gray}{15.33} \\
       PseudoBinaural~\citeyearpar{xu2021visually}\tnote{*} & 7.23 & 2.81 & 0.65 & 1.85 & 6.39 & 4.00 & 10.36 & 12.91 \\
       Ours & \textbf{4.93} & \textbf{1.44} & \textbf{0.58} & \textbf{2.23} & \textbf{3.07} & \textbf{2.45} & \textbf{6.99} & \textbf{8.16} \\ 
       \midrule
       Ours w/o text & 6.77 & 2.54 & 0.63 & 2.00 & 5.87 & 4.03 & 9.25 & 11.40 \\
       Ours w/o Flipper & 5.08 & 1.72 & 0.61 & 2.15 & 4.14 & 2.89 & 8.63 & 10.03 \\
       \bottomrule
    \end{tabular}
    
    \begin{tablenotes}    
        \footnotesize               
        \item[*] indicates that we re-train it on the SpatialTAS dataset.
    \end{tablenotes}            
    \end{threeparttable}       
    
    \caption{\textbf{Results on the testing set of SpatialTAS.} \textcolor{gray}{Mono-Mono} refers to duplicating the mono audio. Our model demonstrates strong performance in both Generation Metrics for audio quality and Understanding Metrics for spatial semantic correctness. Additionally, we present ablation results without text conditions and the flipped-channel audio augmentation module.}
    \label{tab:main_results}
\end{table*}

\noindent \textbf{Revisiting FAIR-Play and YouTube-Binaural Dataset.} The FAIR-Play Dataset~\citep{gao20192} contains 1,871 ten-second video clips accompanied by binaural audio recordings, totaling 5.2 hours of content, primarily focused on musical instrument sounds. To evaluate our model further, we also use the audio from the YouTube-Binaural Dataset~\citep{garg2023visually}, which includes 426 corresponding 360° videos. This dataset is sourced from the YT-Clean dataset~\citep{morgado2018self}, featuring in-the-wild 360° YouTube videos collected using spatial audio-related queries, with limited superimposed sources like room conversations and individuals playing instruments. For fair comparison, we extract one frame from each video and generate text prompts describing the locations of each sound source. Using GPT-4o~\citep{hurst2024gpt}, we set task parameters related to the Field Of View (FOV) and the receiver's position, instructing it to generate captions based on the frames. More details about the caption generation process can be found in \Cref{sec:img-cap}.

\subsection{Evaluation Metrics} 
During evaluation, we use both generation metrics and understanding metrics to assess the generated binaural audio. The generation quality metrics include \textit{Fréchet Distance} (FD), \textit{Fréchet Audio Distance} (FAD), \textit{Kullback-Leibler Divergence} (KL), and \textit{Inception Score} (IS). We also compare our model with previous non-generative models using \textit{STFT Distance} (STFT) and \textit{Envelope Distance} (ENV). The understanding metrics comprise \textit{Direction of Arrival} (DOA) and \textit{Distance Estimation} (DE) for perception-related questions, as well as \textit{Direction} and \textit{Distance} for reasoning questions. More details about these metrics are provided in \Cref{sec:metrics_details}.

\section{Results}

We first present the experimental results on the test set of the proposed SpatialTAS dataset, using both generation and understanding metrics. Next, we report results from the real-recorded FAIR-Play dataset~\citep{gao20192} and the 360$^{\circ}$ Youtube-Binaural dataset~\citep{garg2023visually}, with the corresponding image-to-caption text descriptions. We then conduct ablation studies focused on the effects of separately modifying the direction, distance, or relative position components of the text prompts. Finally, we visualize several generated results alongside their spectrograms, using different kind of text prompts.

\subsection{SpatialTAS Evaluation Results}
The performance of our model is comprehensively evaluated on the testing set of SpatialTAS. The detailed results are presented in Table~\ref{tab:main_results}, where we compare our approach with two baselines: \textit{Mono-Mono} and PseudoBinaural~\citep{xu2021visually}. \textit{Mono-Mono} serves as a baseline to verify whether our model can effectively distinguish between the two channels, achieved by duplicating the same monaural audio to create a two-channel input. PseudoBinaural~\citep{xu2021visually} shares a similar concept with our method in leveraging large-scale pseudo-generated binaural audio for training and demonstrating generalization to real audio. Originally proposed with a U-Net structure and cross-attention mechanism utilizing extracted visual features, we re-train PseudoBinaural on SpatialTAS with the corresponding text descriptions to ensure a fair comparison. 

As detailed in Table~\ref{tab:main_results}, we conduct an extensive comparison of the models on a range of quality metrics that evaluate the overall quality of the generated audio, as well as spatialization metrics that specifically assess the accuracy of spatialization achieved through text-based spatial question-answering. Our model consistently demonstrates superior performance across multiple metrics, particularly in the spatial perception and reasoning tasks, which involve evaluating the generated audio based on questions focusing on "the relative positions between any two sounding sources" and "estimating the relative distance between any two sounding sources". Notably, in the reasoning part of the understanding metrics, we observe a significant performance improvement of 5.80\% and 7.17\% compared to the Mono-Mono baseline. In contrast, the PseudoBinaural approach achieves improvements of only 2.43\% and 2.42\% over Mono-Mono. This observation suggests that PseudoBinaural may lack the necessary capabilities to effectively generate corresponding binaural audio guided by relative position text descriptions. To further analyze the impact of different components in our model, we conduct ablation studies by evaluating models without text-guided descriptions and models trained without the text spatial coherence augmentation (i.e., without the binaural channel flippers). The results clearly demonstrate the significance of both text conditions and the spatial coherence module in achieving superior performance.

\subsection{Real-recorded Binaural Audio Evaluation}
\label{sec:in-the-wild results}

We extend our evaluation to the FAIR-Play Dataset and the 360$^{\circ}$ Youtube-Binaural Dataset, which encompass in-the-wild binaural audio recordings of music and life-like sounds. Since these datasets are originally video-based, we generate text descriptions for the locations of each sounding source based on the videos using GPT-4o~\citep{hurst2024gpt}. Notably, we generate different spatial position descriptions according to the extracted frames with varying Field of View (FOV), considering that the extracted frames in the FAIR-Play Dataset are not 360$^{\circ}$ views, while those in the 360$^{\circ}$ Youtube-Binaural Dataset are omnidirectional views. This approach ensures that our model is evaluated on a diverse set of real-world binaural audio recordings. 

\begin{table}
    \centering
    \tabcolsep=1pt
    \begin{tabular}{lcccc}
       \toprule
       \multirow{2}{*}{Method}  &  \multicolumn{4}{c}{FAIR-Play Dataset} \\
       & STFT$\downarrow$ & ENV$\downarrow$ & WAV$\downarrow$ & SNR$\uparrow$ \\
       \midrule
       \textcolor{gray}{Mono-Mono} & \textcolor{gray}{1.155} & \textcolor{gray}{0.153} & \textcolor{gray}{7.666} & \textcolor{gray}{5.735} \\
       \midrule
       L2BNet~\citeyearpar{rachavarapu2021localize} & 1.028 & 0.148 & - & - \\
       Mono2Binaural~\citeyearpar{gao20192} & 0.959 & 0.141 & 6.496 & 6.232 \\
       APNet~\citeyearpar{zhou2020sep} & 0.889 & 0.136 & 5.758 & 6.972 \\
       Sep-binaural~\citeyearpar{zhou2020sep} & 0.879 & 0.135 & 6.526 & 6.422 \\
       Main-net~\citeyearpar{zhang2021multi} & 0.867 & 0.135 & 5.750 & 6.985 \\
       Complete-net~\citeyearpar{zhang2021multi} & 0.856 & 0.134 & 5.787 & 6.959 \\
       AVSN~\citeyearpar{liu2024visually} & 0.849 & 0.133 & - & - \\
       Cyclic~\citeyearpar{li2024cyclic} & 0.787 & 0.128 & 5.244 & 7.546 \\
       \midrule
       TAS~\citeyearpar{li2024tas} & 0.914 & 0.137 & 6.092 & 6.771 \\
       Ours & \textbf{0.773} & \textbf{0.126} & \textbf{5.019} & \textbf{7.966} \\ 
       \bottomrule
    \end{tabular}
    \caption{\textbf{Results on the FAIR-Play Dataset.} Our model performs well in real-world scenarios with diverse musical sound sources and outperforms visually guided models, underscoring the importance of text prompts.}
    \label{tab:fair-play}
\end{table}

As comprehensively presented in Table~\ref{tab:fair-play}, we conduct an extensive comparison of our model with other visual-guided and text-guided methods. Our model consistently outperforms the others across almost all metrics. It is noticing that TAS~\cite{li2024tas} exhibits inferior performance compared to previous visual-guided methods. In contrast, our method surpasses these visual-guided methods. This observation suggests that utilizing more flexible text descriptions for the location of sounding sources, encompassing both 3D spatial position descriptions and relative position descriptions, provides the model with more generalized guidance for audio spatialization. Furthermore, as illustrated in Table~\ref{tab:youtube-binaural}, we demonstrate performance improvements on the 360$^{\circ}$ Youtube-Binaural Dataset, showcasing the generalization capabilities of our model to in-the-wild scenarios.

\begin{table}
    \centering
    \begin{tabular}{lcc}
       \toprule
       \multirow{2}{*}{Method}  &  \multicolumn{2}{c}{360$^{\circ}$ Youtube-Binaural} \\
       & STFT$\downarrow$ & ENV$\downarrow$ \\
       \midrule
       \textcolor{gray}{Mono-Mono} & \textcolor{gray}{4.715} & \textcolor{gray}{0.261} \\
       Audio-only & 3.129 & 0.213 \\
       Mono2Binaural~\citeyearpar{gao20192} & 2.892 & 0.208 \\
       APNet~\citeyearpar{liu2024visually} & 2.733 & 0.204 \\
       SimBinaural~\citeyearpar{garg2023visually} & 2.544 & 0.196 \\      
       \midrule
       Ours & \textbf{2.471} & \textbf{0.188} \\
       \bottomrule
    \end{tabular}
    \caption{\textbf{Results on the 360$^{\circ}$ Youtube-Binaural Dataset.} The results indicate that our model easily extends to various types of real-recorded sounds, including speech and diverse natural sounds.}
    \label{tab:youtube-binaural}
\end{table}

\subsection{Ablations for Text Prompts}
\begin{figure}[t]
    \centering
    \includegraphics[width=0.5\textwidth]{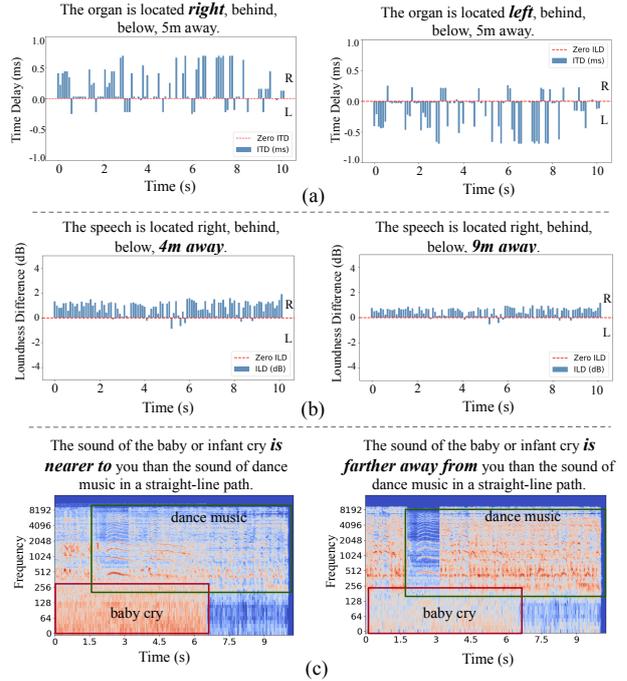}
    \caption{\textbf{Ablations for text prompts.} We systematically alter the direction, distance, and relative position in the text prompts, and present the differences observed before and after these changes.}
    \label{fig:ablation} 
\end{figure}
As illustrated in \Cref{fig:ablation}, we present the results of our ablation studies, focusing on how changes in text prompts related to direction, distance, and multi-source relative positions affect sound localization. \Cref{fig:ablation}(a) demonstrates the impact of changing the directional component of the text prompt from "right" to "left". This adjustment enables us to evaluate the Interaural Time Difference (ITD), which measures the time delay for sound to reach each ear. The goal of estimating the ITD is to ascertain the difference in arrival times of a sound at two microphones. Our results indicate that modifying the directional aspect effectively localizes the sound to the specified direction. \Cref{fig:ablation}(b) illustrates the Interaural Level Difference (ILD) when the distance of the sound source is changed from "4m away" to "9m away". The ILD refers to the difference in sound pressure levels reaching each ear. We observe that altering the distance results in a lower ILD, demonstrating how distance affects perceived loudness. \Cref{fig:ablation}(c) represents the differences in spectrograms when changing the relative position from "is nearer to" to "is farther away from". Given the significant frequency differences between the sounds of a baby crying and dance music, we can analyze the changes in the directional spectrogram by examining variations in energy levels. The sound of a baby crying primarily occupies the lower left section of the spectrogram, while dance music predominantly occupies the upper region. This results in a noticeable change in energy: the baby cry exhibits a transition from high to low energy, whereas the dance music shows a shift from low to high power. Overall, these findings demonstrate that text prompts can provide more detailed and flexible control over the localization of sound sources.

\subsection{Visualization Results}
\begin{figure}[t]
    \centering
    \includegraphics[width=0.45\textwidth]{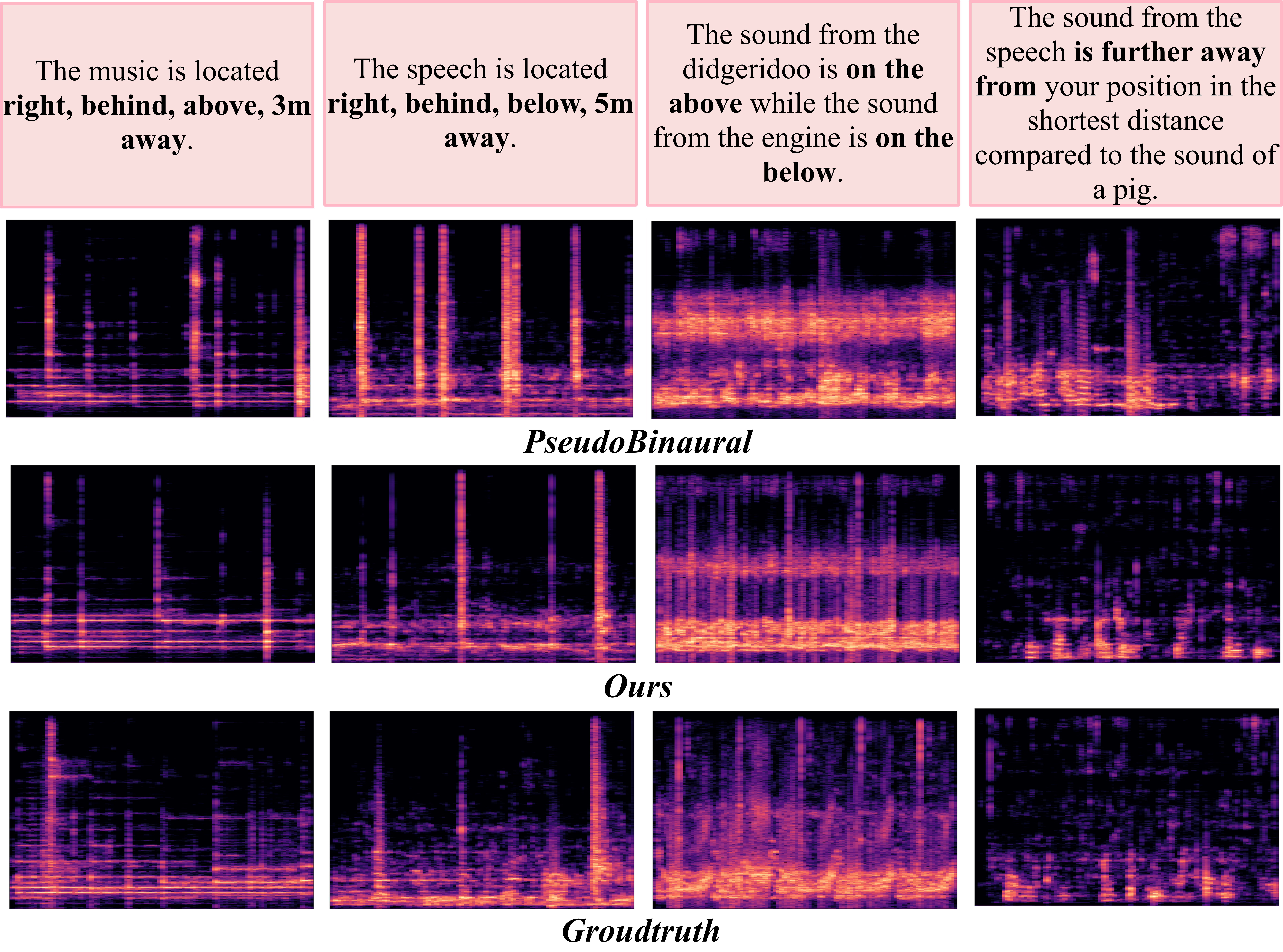}
    \caption{\textbf{Visualization for binaural difference prediction.} We present the binaural difference results using various spatial text prompts, including 3D sound source descriptions and relative position descriptions for music, speech, and natural sounds.}
    \label{fig:visualization} 
\end{figure}
As illustrated in \Cref{fig:visualization}, we present several generation results from the test set of the SpatialTAS dataset. First, we display text prompts that provide detailed descriptions of single-object locations for music and speech. Next, we showcase flexible text prompts designed for multi-object relative location descriptions for a broader range of natural sounds. The results indicate that our method generates audio with a more natural distribution that closely aligns with the ground truth compared to PseudoBinaural.

\section{Conclusion}
We propose a Text-guided Audio Spatialization (TAS) framework, providing a more flexible and interactive control to map monaural audios to binaural ones. We especially train the latent diffusion model on large-scale simulated datasets and can perform well on real-recorded datasets. We evaluate the binaural audio quality from generation metrics and spatial coherence through spatial audio reasoning with LLM. Results show that we can generate binaural audios with both high-quality and semantic consistency in spatial locations. 

\section*{Limitations}
Our model does not account for changes in the location of each sounding object. For instance, a car approaching the listener would produce a change in perceived distance from far to near. Additionally, due to data limitations, our model currently relies solely on text modality to guide audio spatialization. We do not incorporate both text and image modalities, or even motion cues from videos as conditioning factors. 

 \section*{Acknowledgments}
 This work was supported by the National Natural Science Foundation of China (Grant No. 62402211) and the Natural Science Foundation of Jiangsu Province (Grant No. BK20241248). Jie Liu is the corresponding author.

\bibliography{main}

\begin{thebibliography}{42}
\providecommand{\natexlab}[1]{#1}

\bibitem[{Chung et~al.(2024)Chung, Hou, Longpre, Zoph, Tay, Fedus, Li, Wang, Dehghani, Brahma et~al.}]{chung2024scaling}
Hyung~Won Chung, Le~Hou, Shayne Longpre, Barret Zoph, Yi~Tay, William Fedus, Yunxuan Li, Xuezhi Wang, Mostafa Dehghani, Siddhartha Brahma, et~al. 2024.
\newblock Scaling instruction-finetuned language models.
\newblock \emph{Journal of Machine Learning Research}, 25(70):1--53.

\bibitem[{Dagli et~al.(2024)Dagli, Prakash, Wu, and Khosravani}]{dagli2024see}
Rishit Dagli, Shivesh Prakash, Robert Wu, and Houman Khosravani. 2024.
\newblock See-2-sound: Zero-shot spatial environment-to-spatial sound.
\newblock \emph{arXiv preprint arXiv:2406.06612}.

\bibitem[{Dubey et~al.(2024)Dubey, Jauhri, Pandey, Kadian, Al-Dahle, Letman, Mathur, Schelten, Yang, Fan et~al.}]{dubey2024llama}
Abhimanyu Dubey, Abhinav Jauhri, Abhinav Pandey, Abhishek Kadian, Ahmad Al-Dahle, Aiesha Letman, Akhil Mathur, Alan Schelten, Amy Yang, Angela Fan, et~al. 2024.
\newblock The llama 3 herd of models.
\newblock \emph{arXiv preprint arXiv:2407.21783}.

\bibitem[{Elizalde et~al.(2023)Elizalde, Deshmukh, Al~Ismail, and Wang}]{elizalde2023clap}
Benjamin Elizalde, Soham Deshmukh, Mahmoud Al~Ismail, and Huaming Wang. 2023.
\newblock Clap learning audio concepts from natural language supervision.
\newblock In \emph{ICASSP 2023-2023 IEEE International Conference on Acoustics, Speech and Signal Processing (ICASSP)}, pages 1--5. IEEE.

\bibitem[{Evans et~al.(2024{\natexlab{a}})Evans, Parker, Carr, Zukowski, Taylor, and Pons}]{evans2024long}
Zach Evans, Julian~D Parker, CJ~Carr, Zack Zukowski, Josiah Taylor, and Jordi Pons. 2024{\natexlab{a}}.
\newblock Long-form music generation with latent diffusion.
\newblock \emph{arXiv preprint arXiv:2404.10301}.

\bibitem[{Evans et~al.(2024{\natexlab{b}})Evans, Parker, Carr, Zukowski, Taylor, and Pons}]{evans2024stable}
Zach Evans, Julian~D Parker, CJ~Carr, Zack Zukowski, Josiah Taylor, and Jordi Pons. 2024{\natexlab{b}}.
\newblock Stable audio open.
\newblock \emph{arXiv preprint arXiv:2407.14358}.

\bibitem[{Fonseca et~al.(2021)Fonseca, Favory, Pons, Font, and Serra}]{fonseca2021fsd50k}
Eduardo Fonseca, Xavier Favory, Jordi Pons, Frederic Font, and Xavier Serra. 2021.
\newblock Fsd50k: an open dataset of human-labeled sound events.
\newblock \emph{IEEE/ACM Transactions on Audio, Speech, and Language Processing}, 30:829--852.

\bibitem[{Gao and Grauman(2019)}]{gao20192}
Ruohan Gao and Kristen Grauman. 2019.
\newblock 2.5 d visual sound.
\newblock In \emph{Proceedings of the IEEE/CVF Conference on Computer Vision and Pattern Recognition}, pages 324--333.

\bibitem[{Garg et~al.(2023)Garg, Gao, and Grauman}]{garg2023visually}
Rishabh Garg, Ruohan Gao, and Kristen Grauman. 2023.
\newblock Visually-guided audio spatialization in video with geometry-aware multi-task learning.
\newblock \emph{International Journal of Computer Vision}, 131(10):2723--2737.

\bibitem[{Gemmeke et~al.(2017)Gemmeke, Ellis, Freedman, Jansen, Lawrence, Moore, Plakal, and Ritter}]{gemmeke2017audio}
Jort~F Gemmeke, Daniel~PW Ellis, Dylan Freedman, Aren Jansen, Wade Lawrence, R~Channing Moore, Manoj Plakal, and Marvin Ritter. 2017.
\newblock Audio set: An ontology and human-labeled dataset for audio events.
\newblock In \emph{2017 IEEE international conference on acoustics, speech and signal processing (ICASSP)}, pages 776--780. IEEE.

\bibitem[{Hershey et~al.(2017)Hershey, Chaudhuri, Ellis, Gemmeke, Jansen, Moore, Plakal, Platt, Saurous, Seybold et~al.}]{hershey2017cnn}
Shawn Hershey, Sourish Chaudhuri, Daniel~PW Ellis, Jort~F Gemmeke, Aren Jansen, R~Channing Moore, Manoj Plakal, Devin Platt, Rif~A Saurous, Bryan Seybold, et~al. 2017.
\newblock Cnn architectures for large-scale audio classification.
\newblock In \emph{2017 ieee international conference on acoustics, speech and signal processing (icassp)}, pages 131--135. IEEE.

\bibitem[{Ho and Salimans(2022)}]{ho2022classifier}
Jonathan Ho and Tim Salimans. 2022.
\newblock Classifier-free diffusion guidance.
\newblock \emph{arXiv preprint arXiv:2207.12598}.

\bibitem[{Hurst et~al.(2024)Hurst, Lerer, Goucher, Perelman, Ramesh, Clark, Ostrow, Welihinda, Hayes, Radford et~al.}]{hurst2024gpt}
Aaron Hurst, Adam Lerer, Adam~P Goucher, Adam Perelman, Aditya Ramesh, Aidan Clark, AJ~Ostrow, Akila Welihinda, Alan Hayes, Alec Radford, et~al. 2024.
\newblock Gpt-4o system card.
\newblock \emph{arXiv preprint arXiv:2410.21276}.

\bibitem[{Kim et~al.(2019{\natexlab{a}})Kim, Kim, Lee, and Kim}]{kim2019audiocaps}
Chris~Dongjoo Kim, Byeongchang Kim, Hyunmin Lee, and Gunhee Kim. 2019{\natexlab{a}}.
\newblock Audiocaps: Generating captions for audios in the wild.
\newblock In \emph{Proceedings of the 2019 Conference of the North American Chapter of the Association for Computational Linguistics: Human Language Technologies, Volume 1 (Long and Short Papers)}, pages 119--132.

\bibitem[{Kim et~al.(2019{\natexlab{b}})Kim, Remaggi, Jackson, and Hilton}]{kim2019immersive}
Hansung Kim, Luca Remaggi, Philip~JB Jackson, and Adrian Hilton. 2019{\natexlab{b}}.
\newblock Immersive spatial audio reproduction for vr/ar using room acoustic modelling from 360 images.
\newblock In \emph{2019 IEEE Conference on Virtual Reality and 3D User Interfaces (VR)}, pages 120--126. IEEE.

\bibitem[{Kingma(2013)}]{kingma2013auto}
Diederik~P Kingma. 2013.
\newblock Auto-encoding variational bayes.
\newblock \emph{arXiv preprint arXiv:1312.6114}.

\bibitem[{Kong et~al.(2020{\natexlab{a}})Kong, Kim, and Bae}]{kong2020hifi}
Jungil Kong, Jaehyeon Kim, and Jaekyoung Bae. 2020{\natexlab{a}}.
\newblock Hifi-gan: Generative adversarial networks for efficient and high fidelity speech synthesis.
\newblock \emph{Advances in neural information processing systems}, 33:17022--17033.

\bibitem[{Kong et~al.(2020{\natexlab{b}})Kong, Cao, Iqbal, Wang, Wang, and Plumbley}]{kong2020panns}
Qiuqiang Kong, Yin Cao, Turab Iqbal, Yuxuan Wang, Wenwu Wang, and Mark~D Plumbley. 2020{\natexlab{b}}.
\newblock Panns: Large-scale pretrained audio neural networks for audio pattern recognition.
\newblock \emph{IEEE/ACM Transactions on Audio, Speech, and Language Processing}, 28:2880--2894.

\bibitem[{Lee et~al.(2023)Lee, Yeon, Nam, and Chung}]{lee2023voiceldmtexttospeechenvironmentalcontext}
Yeonghyeon Lee, Inmo Yeon, Juhan Nam, and Joon~Son Chung. 2023.
\newblock \href {https://arxiv.org/abs/2309.13664} {Voiceldm: Text-to-speech with environmental context}.
\newblock \emph{Preprint}, arXiv:2309.13664.

\bibitem[{Li et~al.(2018)Li, Langlois, and Zheng}]{li2018scene}
Dingzeyu Li, Timothy~R Langlois, and Changxi Zheng. 2018.
\newblock Scene-aware audio for 360 videos.
\newblock \emph{ACM Transactions on Graphics (TOG)}, 37(4):1--12.

\bibitem[{Li et~al.(2024{\natexlab{a}})Li, Zhao, and Yuan}]{li2024cyclic}
Zhaojian Li, Bin Zhao, and Yuan Yuan. 2024{\natexlab{a}}.
\newblock Cyclic learning for binaural audio generation and localization.
\newblock In \emph{Proceedings of the IEEE/CVF Conference on Computer Vision and Pattern Recognition}, pages 26669--26678.

\bibitem[{Li et~al.(2024{\natexlab{b}})Li, Zhao, and Yuan}]{li2024tas}
Zhaojian Li, Bin Zhao, and Yuan Yuan. 2024{\natexlab{b}}.
\newblock Tas: Personalized text-guided audio spatialization.
\newblock In \emph{ACM Multimedia 2024}.

\bibitem[{Liu et~al.(2023)Liu, Chen, Yuan, Mei, Liu, Mandic, Wang, and Plumbley}]{liu2023audioldm}
Haohe Liu, Zehua Chen, Yi~Yuan, Xinhao Mei, Xubo Liu, Danilo Mandic, Wenwu Wang, and Mark~D Plumbley. 2023.
\newblock Audioldm: Text-to-audio generation with latent diffusion models.
\newblock \emph{arXiv preprint arXiv:2301.12503}.

\bibitem[{Liu et~al.(2024{\natexlab{a}})Liu, Yuan, Liu, Mei, Kong, Tian, Wang, Wang, Wang, and Plumbley}]{liu2024audioldm}
Haohe Liu, Yi~Yuan, Xubo Liu, Xinhao Mei, Qiuqiang Kong, Qiao Tian, Yuping Wang, Wenwu Wang, Yuxuan Wang, and Mark~D Plumbley. 2024{\natexlab{a}}.
\newblock Audioldm 2: Learning holistic audio generation with self-supervised pretraining.
\newblock \emph{IEEE/ACM Transactions on Audio, Speech, and Language Processing}.

\bibitem[{Liu et~al.(2024{\natexlab{b}})Liu, Wang, Qian, and Xie}]{liu2024visually}
Miao Liu, Jing Wang, Xinyuan Qian, and Xiang Xie. 2024{\natexlab{b}}.
\newblock Visually guided binaural audio generation with cross-modal consistency.
\newblock In \emph{ICASSP 2024-2024 IEEE International Conference on Acoustics, Speech and Signal Processing (ICASSP)}, pages 7980--7984. IEEE.

\bibitem[{Liu et~al.(2024{\natexlab{c}})Liu, Chen, Bai, Liang, Li, Gao, and Lin}]{liu2024aligning}
Yang Liu, Weixing Chen, Yongjie Bai, Xiaodan Liang, Guanbin Li, Wen Gao, and Liang Lin. 2024{\natexlab{c}}.
\newblock Aligning cyber space with physical world: A comprehensive survey on embodied ai.
\newblock \emph{arXiv preprint arXiv:2407.06886}.

\bibitem[{Loshchilov(2017)}]{loshchilov2017decoupled}
I~Loshchilov. 2017.
\newblock Decoupled weight decay regularization.
\newblock \emph{arXiv preprint arXiv:1711.05101}.

\bibitem[{Morgado et~al.(2018)Morgado, Nvasconcelos, Langlois, and Wang}]{morgado2018self}
Pedro Morgado, Nuno Nvasconcelos, Timothy Langlois, and Oliver Wang. 2018.
\newblock Self-supervised generation of spatial audio for 360 video.
\newblock \emph{Advances in neural information processing systems}, 31.

\bibitem[{Parida et~al.(2022)Parida, Srivastava, and Sharma}]{parida2022beyond}
Kranti~Kumar Parida, Siddharth Srivastava, and Gaurav Sharma. 2022.
\newblock Beyond mono to binaural: Generating binaural audio from mono audio with depth and cross modal attention.
\newblock In \emph{Proceedings of the IEEE/CVF Winter Conference on Applications of Computer Vision}, pages 3347--3356.

\bibitem[{Rachavarapu et~al.(2021)Rachavarapu, Sundaresha, Rajagopalan et~al.}]{rachavarapu2021localize}
Kranthi~Kumar Rachavarapu, Vignesh Sundaresha, AN~Rajagopalan, et~al. 2021.
\newblock Localize to binauralize: Audio spatialization from visual sound source localization.
\newblock In \emph{Proceedings of the IEEE/CVF International Conference on Computer Vision}, pages 1930--1939.

\bibitem[{Raffel et~al.(2020)Raffel, Shazeer, Roberts, Lee, Narang, Matena, Zhou, Li, and Liu}]{raffel2020exploring}
Colin Raffel, Noam Shazeer, Adam Roberts, Katherine Lee, Sharan Narang, Michael Matena, Yanqi Zhou, Wei Li, and Peter~J Liu. 2020.
\newblock Exploring the limits of transfer learning with a unified text-to-text transformer.
\newblock \emph{Journal of machine learning research}, 21(140):1--67.

\bibitem[{Rombach et~al.(2022)Rombach, Blattmann, Lorenz, Esser, and Ommer}]{rombach2022high}
Robin Rombach, Andreas Blattmann, Dominik Lorenz, Patrick Esser, and Bj{\"o}rn Ommer. 2022.
\newblock High-resolution image synthesis with latent diffusion models.
\newblock In \emph{Proceedings of the IEEE/CVF conference on computer vision and pattern recognition}, pages 10684--10695.

\bibitem[{Singh~Kushwaha et~al.(2024)Singh~Kushwaha, Ma, Thomas, Tian, and Bruni}]{singh2024diff}
Saksham Singh~Kushwaha, Jianbo Ma, Mark~RP Thomas, Yapeng Tian, and Avery Bruni. 2024.
\newblock Diff-sage: End-to-end spatial audio generation using diffusion models.
\newblock \emph{arXiv e-prints}, pages arXiv--2410.

\bibitem[{Song et~al.(2020)Song, Meng, and Ermon}]{song2020denoising}
Jiaming Song, Chenlin Meng, and Stefano Ermon. 2020.
\newblock Denoising diffusion implicit models.
\newblock \emph{arXiv preprint arXiv:2010.02502}.

\bibitem[{Sun et~al.(2024)Sun, Cheng, Li, Ye, Liu, Zhang, Xue, and Guo}]{sun2024both}
Peiwen Sun, Sitong Cheng, Xiangtai Li, Zhen Ye, Huadai Liu, Honggang Zhang, Wei Xue, and Yike Guo. 2024.
\newblock Both ears wide open: Towards language-driven spatial audio generation.
\newblock \emph{arXiv preprint arXiv:2410.10676}.

\bibitem[{Vyas et~al.(2023)Vyas, Shi, Le, Tjandra, Wu, Guo, Zhang, Zhang, Adkins, Ngan et~al.}]{vyas2023audiobox}
Apoorv Vyas, Bowen Shi, Matthew Le, Andros Tjandra, Yi-Chiao Wu, Baishan Guo, Jiemin Zhang, Xinyue Zhang, Robert Adkins, William Ngan, et~al. 2023.
\newblock Audiobox: Unified audio generation with natural language prompts.
\newblock \emph{arXiv preprint arXiv:2312.15821}.

\bibitem[{Xu et~al.(2024)Xu, Markovic, Sandakly, Keebler, Krenn, and Richard}]{xu2024sounding}
Xudong Xu, Dejan Markovic, Jacob Sandakly, Todd Keebler, Steven Krenn, and Alexander Richard. 2024.
\newblock Sounding bodies: modeling 3d spatial sound of humans using body pose and audio.
\newblock \emph{Advances in Neural Information Processing Systems}, 36.

\bibitem[{Xu et~al.(2021)Xu, Zhou, Liu, Dai, Wang, and Lin}]{xu2021visually}
Xudong Xu, Hang Zhou, Ziwei Liu, Bo~Dai, Xiaogang Wang, and Dahua Lin. 2021.
\newblock Visually informed binaural audio generation without binaural audios.
\newblock In \emph{Proceedings of the IEEE/CVF Conference on Computer Vision and Pattern Recognition}, pages 15485--15494.

\bibitem[{Yang et~al.(2023)Yang, Tian, Tan, Huang, Liu, Chang, Shi, Zhao, Bian, Wu, Zhao, Watanabe, and Meng}]{yang2023uniaudioaudiofoundationmodel}
Dongchao Yang, Jinchuan Tian, Xu~Tan, Rongjie Huang, Songxiang Liu, Xuankai Chang, Jiatong Shi, Sheng Zhao, Jiang Bian, Xixin Wu, Zhou Zhao, Shinji Watanabe, and Helen Meng. 2023.
\newblock \href {https://arxiv.org/abs/2310.00704} {Uniaudio: An audio foundation model toward universal audio generation}.
\newblock \emph{Preprint}, arXiv:2310.00704.

\bibitem[{Zhang and Shao(2021)}]{zhang2021multi}
Wen Zhang and Jie Shao. 2021.
\newblock Multi-attention audio-visual fusion network for audio spatialization.
\newblock In \emph{Proceedings of the 2021 International Conference on Multimedia Retrieval}, pages 394--401.

\bibitem[{Zheng et~al.(2024)Zheng, Peng, Ma, Chen, Choi, and Harwath}]{zheng2024bat}
Zhisheng Zheng, Puyuan Peng, Ziyang Ma, Xie Chen, Eunsol Choi, and David Harwath. 2024.
\newblock Bat: Learning to reason about spatial sounds with large language models.
\newblock \emph{arXiv preprint arXiv:2402.01591}.

\bibitem[{Zhou et~al.(2020)Zhou, Xu, Lin, Wang, and Liu}]{zhou2020sep}
Hang Zhou, Xudong Xu, Dahua Lin, Xiaogang Wang, and Ziwei Liu. 2020.
\newblock Sep-stereo: Visually guided stereophonic audio generation by associating source separation.
\newblock In \emph{Computer Vision--ECCV 2020: 16th European Conference, Glasgow, UK, August 23--28, 2020, Proceedings, Part XII 16}, pages 52--69. Springer.

\end{thebibliography}

\clearpage
\appendix

\section{Image Caption Engineering}
\label{sec:img-cap}
\begin{figure}[ht]
    \centering
    \includegraphics[width=0.45\textwidth]{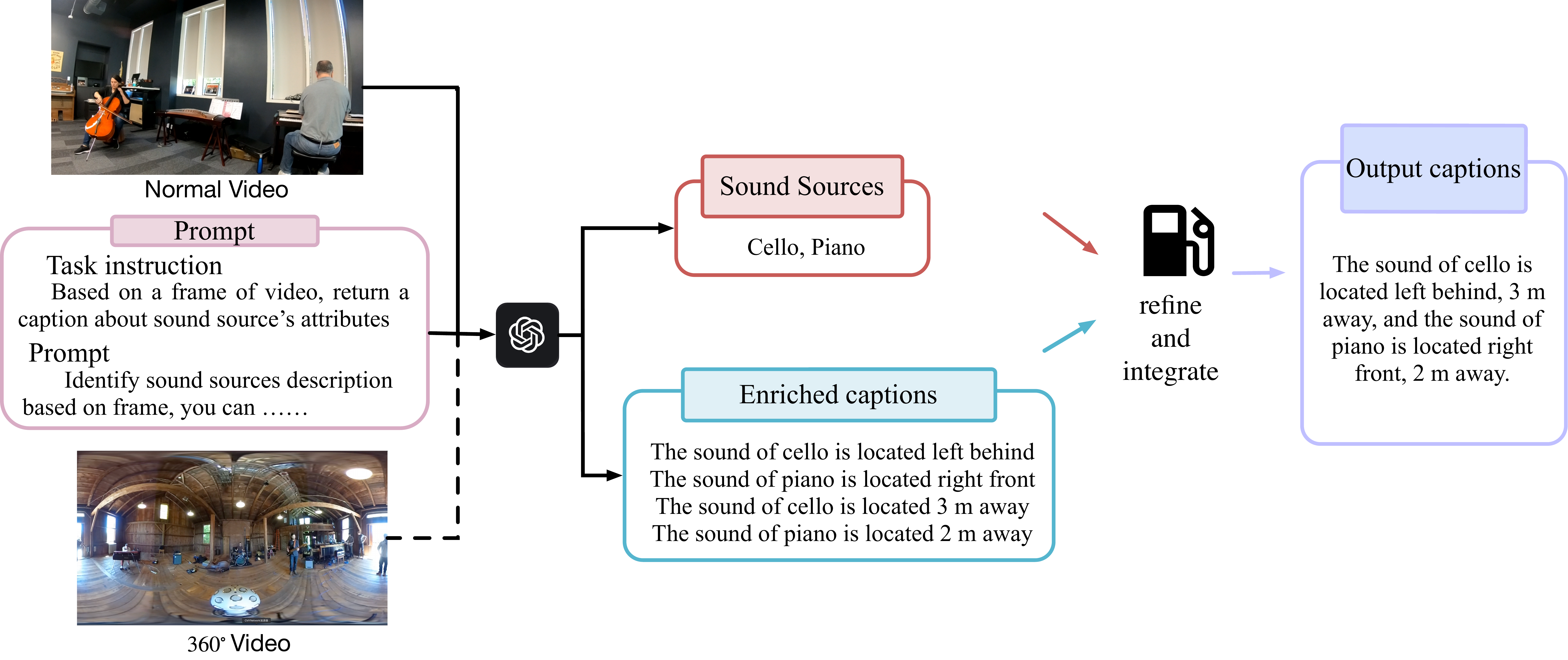}
    \caption{The caption engineering and an example.}
    \label{fig:caption} 
\end{figure}

We extract the required sound sources of the video frame and its corresponding description of the orientation and distance, which is summarized into a caption using large language model(LLM). Prompting allows a pre-trained model to adapt to different tasks via different prompts without modifying any parameters. LLMs like GPT-4o have shown strong zero-shot and few-shot ability via prompting. Prompting has been successful for a variety of natural language tasks, hence we design prompt for GPT-4o for sound source detection and attribute inference in images. We provide an image of video and a list of detected sound sources. Then we require sound objects with attributes (relative orientation and distance from the lens). The prompt-guided caption complies with (1) accurately detect the sound source object (2) describe the required attributes as general captions do, and (3) provide auxiliary information in the caption if necessary. \Cref{fig:caption} illustrates the GPT-4o prompt we use for image caption engineering. 

\section{Evaluation Details}
\label{sec:metrics_details}
FD and FAD assess the distribution similarity between real and generated audio using different classifiers, with FAD employing VGGish~\citep{hershey2017cnn} and FD using PANNs~\citep{kong2020panns}. KL quantifies distribution similarity, while IS evaluates the quality and diversity of the generated audio. Additionally, we compare our model with previous non-generative models using \textit{STFT Distance} (STFT) and \textit{Envelope Distance} (ENV). STFT is calculated as the Euclidean distance between the ground-truth and predicted complex spectrograms, scaled to represent raw audio energy levels. ENV involves computing the envelope of both ground-truth and predicted waveforms, as direct waveform comparisons may not capture perceptual similarity effectively.

\section{QA Pairs for Spatial Audio Reasoning}
\label{sec:qa-type}
As shown in \Cref{tab:data_qa}, the questions can be categorized into spatial perception and spatial reasoning types. The perception questions primarily focus on Direction of Arrival (DOA) and Distance Estimation (DE), addressing the direction and distance descriptions for each sound source. In contrast, the reasoning questions involve the relative direction and distance between any two sound sources.

\begin{table*}[ht]
    \centering
    \footnotesize
    \setlength\tabcolsep{1.5pt}
    \caption{\textbf{QA pairs used the spatial llm reasoning task.} The first four types focus on perception, while the last emphasizes reasoning. DP: Distance Prediction; DOA: Direction-of-Arrival. Numbers (e.g., 139K, 15.9\%) indicate the QA sample count and their percentages in the dataset.}
    \label{tab:data_qa}
    \begin{tabular}{lcl}
    \toprule
       \textbf{Spatialization} & \textbf{QA Type} & \textbf{Example} \\
    \midrule
        \multirow{6}{*}{\textbf{Perception}} & \multirow{6}{*}{\shortstack[l]{DOA \& \\ DER} } & \textbf{Q:} (In single sound source.) How would you describe the location of the music's sound in terms of \\
        & & \quad \ direction and distance? \\
        & & \textbf{A:} right, behind, below; 4m \\
        & & \textbf{Q:} (In double sound source.) At what distance and in which direction, is the writing's sound originating? \\
        & & \textbf{A:} left, behind, above; 2.5m \\ 
    \midrule 
        \multirow{10}{*}{\textbf{Reasoning}} & \multirow{10}{*}{\shortstack[l]{Direction \& \\ Distance}} &  \textbf{Q:} Measuring the shortest path in a straight line, is the sound of camera more distant from you \\
        & & \quad \ than the sound of music? / \textbf{A:} Yes \\
        & & \textbf{Q:} Is the sound of bird flight, flapping wings further from you than music when considering \\
        & & \quad \ the direct paths? / \textbf{A:} Yes \\
        & & \textbf{Q:} Can you estimate the distance from the sound of the speech to the sound of the drawer open or close? \\
        && \textbf{A:} 1.5m \\
        & & \textbf{Q:} What is the sound on the below side of the sound of the wind instrument, woodwind instrument? \\ 
        & & \textbf{A:} slap, smack \\
        & & \textbf{Q:} Could you determine whether the breaking's sound is to the left or right of the music's sound? \\
        & & \textbf{A:} left \\
    \bottomrule
    \end{tabular}
\end{table*}

\section{Discussion about failure cases}
\noindent\textbf{Case 1} When two sources have similar characteristics with similar energy distributions in spectrogram, the generated results lead to distortion for the text embeddings may map to the same spectrogram part. In the future, we will specifically apply spectrogram-similar sound sources and spectrogram-different sound sources for targeted analysis.

\noindent\textbf{Case 2} Incorrect text-embedding to audio mapping can result in unwanted sounds, especially in speech, which presents more stringent requirements compared to music and natural sounds. To address this issue, we will curate a diverse and representative dataset, employ advanced embedding techniques to capture nuanced differences, implement regularization methods to mitigate overfitting, and apply domain adaptation tailored to specific audio types.
\end{document}